\documentclass[fleqn,twoside]{article}
\usepackage[headings]{espcrc2}
\usepackage{slashed}
\usepackage{cite}

\readRCS
$Id: espcrc2.tex,v 1.2 2004/02/24 11:22:11 spepping Exp $
\usepackage{graphicx}
\usepackage[figuresright]{rotating}

\newcommand{\AmS}{{\protect\the\textfont2
  A\kern-.1667em\lower.5ex\hbox{M}\kern-.125emS}}

\newcommand{\as}{\alpha_{\rm s}}

\def\ep{\epsilon}
\def\z#1{{\zeta_{#1}}}

\def\cf{{C^{}_F}}

\def\ly{\mathrm{L}_y}
\def\s#1#2{\mathrm{S}_{#1,#2}(x)}
\def\li#1{\mathrm{Li}_{#1}(x)}

\newcommand{\brk}{\right. \nonumber \\ && \left.}
\newcommand{\ibrk}{\right. \right. \nonumber \\ && \left. \left.}
\newcommand{\dibrk}{\right. \right. \right. \nonumber \\ && \left. \left. \left.}

\newcommand{\text}{\textstyle}
\newcommand{\bea}{\begin{eqnarray}}
\newcommand{\eea}{\end{eqnarray}}

\def\Ione#1{\mbox{\boldmath I}_{#1}^{(1)}(\ep)}

\title{
Massive particle production to NNLO in QCD
}

\author{G. Chachamis\address[MCSD]{
Institut f\"ur Theoretische Physik und Astrophysik, 
Universit\"at W\"urzburg \\
Am Hubland, D-97074 W\"urzburg, Germany}
        \thanks{This work was supported by the Sofja Kovalevskaja Award of the
Alexander von Humboldt Foundation.}}
       
\runtitle{2-column format camera-ready paper in \LaTeX}
\runauthor{G. Chachamis}

\begin{document}

\begin{abstract}
We discuss the recent derivation of the one-loop 
squared virtual QCD
corrections to the W boson pair production in the 
quark-anti-quark-annihilation channel in the limit where all 
kinematical invariants are large compared to the mass of the W boson.
In particular, we elaborate on the 
combined use of the helicity matrix formalism
with the Mellin-Barnes representations technique.
\vspace{1pc}
\end{abstract}

\maketitle

\section{Introduction}
\label{sec:intro}

The Large Hadron Collider (LHC) is expected to have a huge impact
on particle physics phenomenology.
Most of the processes which will be studied at the LHC need to be
calculated at least to next-to-leading order (NLO) in QCD whereas there
are some for which a theoretical prediction is needed to 
next-next-to-leading order (NNLO).
Electroweak gauge boson pair production falls into the latter category.
One of the reasons is that the 
increase of the centre-of-mass energy at the LHC 
with respect to the Tevatron 
from 1.96 TeV to 14 TeV will result in a huge boost of the available
data.

The importance of  hadronic W-pair production is two-fold. Firstly, it is a
process which allows the measurement of the vector boson trilinear couplings 
and therefore a comparison with the Standard Model (SM) predictions.
Most attempts to model New Physics, such as 
Supersymmetry and Extra-dimensions in all
variations, should be able to
explain any 
deviations by consistently adjusting the anomalous
couplings and/or by 
incorporating decays of new particles into vector boson 
pairs~\cite{tevatron1,tevatron2}. 

Secondly, hadronic W pair 
production is important for investigations of the nature of the
Electroweak symmetry mechanism by contributing the 
dominant background
for the Higgs boson mediated process 
(see Refs.~\cite{Spira:1995rr, Dawson:1990zj,Harlander:2002wh, 
Anastasiou:2002yz, Ravindran:2003um, Catani:2001cr, Davatz:2004zg, 
Anastasiou:2004xq, Anastasiou:2007mz, Grazzini:2008tf,
Bredenstein:2006rh, kauer1, kauer2}),
\bea
p p \rightarrow H \rightarrow W^* W^* 
\rightarrow l {\bar \nu} {\bar l}' \nu' \, , \nonumber
\eea
in the Higgs mass range between 
$140 \, \mathrm{GeV} <  \mathrm{M}_{\mathrm{H}} 
<  180 \, \mathrm{GeV}$~\cite{dittmardreiner}. 

The interest in hadronic W pair production is well displayed by
the fact that the Born cross 
section was calculated some thirty years ago~\cite{brown}.
The NLO QCD corrections were computed in the 90's and seen to contribute
a 30\%~\cite{ohn,fri,dixon1,dixon2,campbell}.  
Next,
soft gluon resummation
effects were considered in Ref.~\cite{grazzini:2006}
whereas massless fermion-boson scattering was studied at NNLO
in Ref.~\cite{Anastasiou:2002zn}.
The first steps towards a complete NNLO study
were taken with the
computation of the NNLO two-loop~\cite{Chachamis:2007cy,
Chachamis:2008yb, Chachamis:2008fx},
as well as the one-loop squared ~\cite{Chachamis:2008xu}
virtual corrections
in a high energy expansion, $M_{W}^2 \ll$ s, t, u.

The methods used in~\cite{Chachamis:2008xu}, differ
somehow from the ones employed in~\cite{Chachamis:2008yb} though,
both  are a continuation of the techniques used before in
\cite{qqTT,ggTT,Czakon:2004wm,Czakon:2006pa,Actis:2007gi}, 
The difference lies mainly in the fact that for
the one-loop squared corrections the
 helicity matrix formalism was additionally used
to reduce the problem to a small set of integrals,
which in turn were treated  with Mellin-Barnes (MB)
representations~\cite{Smirnov:1999gc,Tausk:1999vh}.
The latter were constructed by means of the {\tt
MBrepresentation} package~\cite{MBrepresentation} 
and then analytically continued in
the number of space-time dimensions $D = 4 -2 \epsilon$ using the
{\tt MB} package~\cite{Czakon:2005rk}. 
After the asymptotic expansion in the mass
parameter, contours were closed and integrals finally resummed either with
the help of {\tt XSummer}~\cite{Moch:2005uc} or the {\tt PSLQ}
algorithm~\cite{pslq:1992}. 

Here, we are going to give more details on how the two methods of
the helicity formalism and MB representations were combined for the
derivation of the result in~\cite{Chachamis:2008xu}. We  provide,
as an example,
the coefficient of a certain helicity matrix element for the one-loop
amplitude in the high energy limit. This result is in closed
analytic form expressed through harmonic polylogarithms and
transcendental constants.

\section{The Calculation}

\label{sec:notation}

We shall introduce here part of the notation
used in~\cite{Chachamis:2008xu}.
The charged vector-boson 
production in the leading partonic scattering process
corresponds to
\begin{equation}
\label{eq:qqWW}
q_j(p_1) + {\overline q}_j(p_2) 
\:\:\rightarrow\:\: W^-(p_3,m) + W^+(p_4,m) \, ,
\end{equation}
where $p_i$ denote 
the quark and W momenta, $m$ is the mass of the W boson and
j is a flavour index.
Here we are considering down-type quark scattering.
Energy-momentum conservation implies
\begin{equation}
\label{eq:engmom}
p_1^\mu+p_2^\mu = p_3^\mu+p_4^\mu \, .
\end{equation}
We consider
the scattering amplitude ${\cal M}$ for the process~(\ref{eq:qqWW})
at fixed values of the external parton momenta $p_i$, thus $p_1^2 =
p_2^2 = 0$ and $p_3^2 = p_4^2 = m^2$.
The amplitude ${\cal M}$ may be written as a 
series expansion in the strong coupling $\as$,
\begin{eqnarray}
  \label{eq:Mexp}
  | {\cal M} \rangle
  & = &
  \biggl[
  | {\cal M}^{(0)} \rangle
  + \biggl( {\as \over 2 \pi} \biggr) | {\cal M}^{(1)} \rangle \nonumber \\
 & + & \biggl( {\as \over 2 \pi} \biggr)^2 | {\cal M}^{(2)} \rangle
  + {\cal O}(\as^3)
  \biggr]
\, .
\end{eqnarray}

For convenience, we define the function ${\cal A}(\epsilon, m, s, t, \mu)$
for the squared amplitudes summed over spins and colors as
\begin{eqnarray}
\label{eq:Msqrd}
\overline{\sum_{\rm{spin}, \rm{color}} |{\cal M}({q_j 
+ {\overline q}_j \to  W^+ + W^-}
  )|^2} \nonumber =  \\ 
{\cal A}(\epsilon, m, s, t, \mu)
\, .
\end{eqnarray}
${\cal A}$ is a function of the Mandelstam variables $s$, $t$ and $u$ given by
$s = (p_1+p_2)^2$, $t  = (p_1-p_3)^2 - m^2$ and $u  = (p_1-p_4)^2 - m^2$
and has a perturbative expansion similar to Eq.~(\ref{eq:Mexp})
\begin{eqnarray}
\label{eq:Aexp}
{\cal A}(\epsilon, m, s, t, \mu) & = &
\left[
  {\cal A}^{(0)}
  + \biggl( {\as \over 2 \pi} \biggr) {\cal A}^{(1)} 
\right. \nonumber \\  
 & + & 
\left. \biggl( {\as \over 2 \pi} \biggr)^2 {\cal A}^{(2)}
  + {\cal O}(\as^{3})
\right]
\, .
\end{eqnarray}
In terms of the amplitudes the expansion coefficients in Eq.~(\ref{eq:Aexp})
may be expressed as
\begin{eqnarray}
\label{eq:A4def}
{\cal A}^{(0)} &=&
\langle {\cal M}^{(0)} | {\cal M}^{(0)}\rangle \, , \\
\label{eq:A6def}
{\cal A}^{(1)} &=& \left(
\langle {\cal M}^{(0)} | {\cal M}^{(1)} \rangle + 
\langle {\cal M}^{(1)} | {\cal M}^{(0)} \rangle
\right)\, , \\
\label{eq:A8def}
{\cal A}^{(2)} &=& \left(
\langle {\cal M}^{(1)} | {\cal M}^{(1)} \rangle
+ \langle {\cal M}^{(0)} | {\cal M}^{(2)} \rangle \right. \nonumber \\ 
&+& \left. \langle {\cal M}^{(2)} | {\cal M}^{(0)} \rangle
\right)\, .
\end{eqnarray}

As already mentioned,
in order to compute  $\langle {\cal M}^{(1)} | {\cal M}^{(1)}
\rangle$ we used the helicity matrix formalism,
namely we expressed  the result in terms of helicity
amplitudes, ${\mathcal M}^g(\lambda_1, \lambda_2, s, t)$.
The quark and the anti-quark have opposite helicities in the
centre-of-mass system so one helicity label above, $g = \pm 1$,
suffices. $\lambda_1$ and $\lambda_2$ stand for the
helicities of the W$^+$ and W$^-$ respectively.

Starting from the one-loop amplitude,
the initial expression can be
rearranged as 
\bea
\label{eq:hme1}
| {\cal M}^{(1)} \rangle = \sum_{i,g}  C_i^g (s,t,u,m) {\mathcal M_i^g}\, , 
\eea
where the $C_i$ are coefficients
and ${\mathcal M_i^g}$ are helicity matrix elements and g = $\pm$.
The ten helicity matrix elements  
have been taken as defined in Ref.~\cite{Diener:1997nx} 
(see also~\cite{Denner:1988tv}):
\bea
{\mathcal M}_0^g &=& {\overline v }(p_2) \, 
\slashed{\epsilon_1} (\slashed{p_3} -
\slashed{p_2}) 
\slashed {\epsilon_2}
{\mathcal P}_g \, u(p_1)  \, , \nonumber \\
{\mathcal M}_1^g &=& {\overline v}(p_2) \, \slashed{p_3} {\mathcal P}_g \, u(p_1) \,
\epsilon_1 \cdot \epsilon_2 \, , \nonumber \\
{\mathcal M}_2^g &=& {\overline v}(p_2)\, \slashed{\epsilon_1} {\mathcal P}_g \, u(p_1)
\, \epsilon_2 \cdot p_3 \, , \nonumber \\
{\mathcal M}_3^g &=& -{\overline v}(p_2)\,\slashed{\epsilon_2} {\mathcal P}_g \, u(p_1)
\, \epsilon_1 \cdot p_4 \, , \nonumber \\
{\mathcal M}_4^g &=& {\overline v}(p_2)\, \slashed{\epsilon_1} {\mathcal P}_g \, u(p_1)
\, \epsilon_2 \cdot p_1\, , \nonumber \\
{\mathcal M}_5^g &=& -{\overline v}(p_2)\, \slashed{\epsilon_2} {\mathcal P}_g \, u(p_1)
\, \epsilon_1 \cdot p_2\, ,  \\
{\mathcal M}_6^g &=& {\overline v}(p_2)\, \slashed{p_3} {\mathcal P}_g \, u(p_1) \,
\epsilon_1 \cdot p_2 \, \epsilon_2 \cdot p_1
\, , \nonumber \\
{\mathcal M}_7^g &=& {\overline v}(p_2)\, \slashed{p_3}{\mathcal P}_g \, u(p_1) \,
\epsilon_1 \cdot p_2 \, \epsilon_2 \cdot p_3 \, , \nonumber \\
{\mathcal M}_8^g &=& {\overline v}(p_2)\, \slashed{p_3}{\mathcal P}_g \, u(p_1)\,
\epsilon_1 \cdot p_4 \, \epsilon_2 \cdot p_1 \, , \nonumber \\
{\mathcal M}_9^g &=& {\overline v}(p_2)\, \slashed{p_3} {\mathcal P}_g \, u(p_1)\,
\epsilon_1 \cdot p_4 \, \epsilon_2 \cdot p_3 \, , \nonumber
\eea
where ${\mathcal P}_g = {\mathcal P}_{\pm} = \frac{1 \pm \gamma_5}{2}$.
All colour indices as well 
as the arguments of the polarization vectors,
$\epsilon_1(p_3, \lambda_1)$ and
$\epsilon_2(p_4, \lambda_2)$, have been suppressed.

Let us have a look at the $C_0^{-}(s, t, u, m)$ coefficient of the  
${\mathcal M}_0^-$ matrix element for the one-loop amplitude.
Typically, $C_0^{-}(s, t, u, m)$ is given by 
\bea
\label{eq:hme2}
C_0^{-}(s, t, u, m) = 
\sum_{i}  c_{0,\,i} (s,t,u,m) 
{\mathcal I}_i (s,t,u,m;\mu^2)
\eea
where $c_{0 \,,i} (s,t,u,m)$ are polynomials in the kinematical
variables and ${\mathcal I}_i (s,t,u,m;\mu^2)$ are one-loop
integrals.
For example, one of the one-loop integrals appearing in Eq.~\ref{eq:hme2}
is
\bea
\label{eq:int}
I = \int d^dk \frac{k.p3}{k^2 (k + p1)^2 (k + p4)^2}
\eea
After feeding this into 
the {\tt MBrepresentation} package, one 
gets, from the reduction of the original tensor 
structure into scalar objects,
the following two terms:
\bea
\label{eq:reps1}
I^{\rm{MB}}_1 &=&
\int_{a - i \infty}^{a + i \infty} d z_1 
(-m^2)^{z_1} (-u)^{-1-\ep-z_1} 
\nonumber \\ && \times
\Gamma(1 - \ep) \Gamma(-\ep - z_1) 
\Gamma(-z_1) 
\nonumber \\ && \times
\Gamma(1 + z_1) \Gamma(1 + \ep + z_1)
\nonumber \\ && \times
\left( \Gamma(2 - 2 \ep) \right)^{-1}
\eea
and
\bea
\label{eq:reps2}
I^{\rm{MB}}_2 &=&
\int_{b - i \infty}^{b + i \infty} d z_1 
(-m^2)^{z_1} (-u)^{-1-\ep-z_1} 
\nonumber \\ && \times
\Gamma(- \ep) \Gamma(1 -\ep - z_1) 
\Gamma(-z_1) 
\nonumber \\ && \times
\Gamma(1 + z_1) \Gamma(1 + \ep + z_1)
\nonumber \\ && \times
\left( \Gamma(2 - 2 \ep) \right)^{-1} \, ,
\eea
where of course $I = I^{\rm{MB}}_1 + I^{\rm{MB}}_2$.
One then needs to perform an asymptotic expansion in the mass
parameter and finally resum the MB integrals. This allows
to compute the coefficients $C_i^g$ in a closed
analytic form in the high energy limit.
As an example, we present here the result up to $\mathcal{O}(\ep^2)$
for $C_0^{-}$ ($\cf$ the color factor and $g^2_{\mathrm{WL}}$
the quark-W coupling)
\begin{eqnarray}
\label{eq:M0}
C_0^{-} &=&
\cf \, g^2_{\mathrm{WL}}
\left \{
\frac{1}{\ep^2}
\left[
-\frac{1}{1-x}
\right]
+
\frac{1}{\ep}
\left[
-\frac{3}{2 (1-x)}
\right]
\right. \nonumber \\ && \left.
+
\frac{i \pi}{\ep}
\left[
-\frac{1}{1-x}
\right]
+
\left[
\frac{7 \pi^2}{12 (1-x)}
\ibrk
+\frac{\ly^2}{2
   (1-x)}
+\frac{3 \ly}{2 (1-x)}
-\frac{9}{2 (1-x)}
\right]
\brk
+
i \pi
\left[
\frac{\ly}{1-x}
\right]
+
\ep
\left[
-\frac{\ly^3}{3 (1-x)}
\ibrk
+\frac{1}{4}
   \left(-\frac{10}{x}-\frac{3}{1-x}\right)
   \ly^2
+\left(\frac{9}{2 (1-x)}
\dibrk
-\frac{\pi^2}{2
   (1-x)}\right) \ly+\frac{\pi^2}{8
   (1-x)}+\frac{1}{3} \left(\frac{7
   \z3}{1-x}
\dibrk
-\frac{27}{1-x}\right)+\frac{\s12}{1-x}
\right]
+
i \pi \, \ep
\left[
\frac{\pi^2}{4 (1-x)}
\ibrk
-\frac{\ly^2}{2
   (1-x)}-\frac{\li2}{1-x}-\frac{5 \ly}{x}
\right]
\brk
+ 
\ep^2
\left[
\frac{3 \pi^2}{8 (1-x)}
+\frac{\li2 \pi^2}{2 (1-x)}
\ibrk
-\frac{73 \pi^4}{1440 (1-x)}
+\frac{\ly^4}{8 (1-x)}
\ibrk
+\frac{1}{12}
   \left(\frac{20}{x}
+\frac{3}{1-x}\right) \ly^3
\ibrk
+\left(\frac{5
   \pi^2}{24 (1-x)}-\frac{3}{4}
   \left(\frac{8}{x}+\frac{3}{1-x}\right)\right)
   \ly^2
\ibrk
+\frac{1}{2} \left(\frac{7
   \z3}{1-x}-\frac{36}{1-x}\right)+\left(\frac{1}{8}
   \left(\frac{20}{x}
\right. \dibrk \left.
-\frac{1}{1-x}\right)
   \pi^2+\frac{9}{1-x}\right)
   \ly-\frac{\ly \s12}{1-x}
\ibrk
-\frac{5
   \s12}{x}-\frac{\s13}{1-x}+\frac{\s22}{1-x}
\right]
\brk
+
i \pi \, \ep^2
\left[
\frac{\ly^3}{6 (1-x)}+\frac{5 \ly^2}{2
   x}
\ibrk
+\left(-\frac{\pi^2}{4 (1-x)}-\frac{12}{x}\right)
   \ly+\frac{\li2 \ly}{1-x}
\ibrk
+\frac{7
   \z3}{3 (1-x)}+\frac{5
   \li2}{x}
\ibrk
-\frac{\li3}{1-x}+\frac{\s12}{1-x}
\right]
\right \} \, ,
\end{eqnarray}
where we have defined 
\bea
x = -\frac{t}{s}\,, \quad y = -\frac{u}{s}\,, \quad m_s = \frac{m^2}{s}
\eea
and
\bea
\label{eq:lmly}
{\rm L_m} = {\rm Log}\left( m_s \right )\, , 
\quad {\rm L_y} = {\rm Log} \left ( 1-x\right ) \, .
\eea

Similarly, one can compute the $C_i^g$ coefficients for 
all the helicity matrix elements ${\mathcal M}_i^g$.
The complex parts of the coefficients are given explicitly
as can be seen in Eq.~\ref{eq:M0} which means that
obtaining the complex conjugate
expressions for the tree-level, 
$| {\cal M}^{(0)} \rangle$,
and the one
loop, $| {\cal M}^{(1)} \rangle$,
amplitudes is trivial.
The following step would be the contraction
$\langle {\cal M}^{(1)} | {\cal M}^{(1)}
\rangle$. As an easy test, we have checked 
$\langle {\cal M}^{(0)} | {\cal M}^{(1)}
\rangle$ derived with this method against the results
provided in~\cite{ohn,fri}.

In order to check $\langle {\cal M}^{(1)} | {\cal M}^{(1)}
\rangle$ we have used a more involved test of
the infrared structure
according to  the Catani prediction~\cite{catani}.
In the case of one-loop QCD amplitudes,
their poles in $\ep$ can be  
expressed as a universal combination of the tree amplitude 
and a colour-charge operator $\Ione{}$. 
The generic form of $\Ione{}$ was found by Catani and 
Seymour~\cite{Catani:1996vz}. 

\section{Conclusions}
\label{sec:conclusions}
We have discussed some details of the computation
of the one-loop squared
NNLO QCD virtual corrections for
the process $q {\bar q} \rightarrow W^+ \, W^-$ in the limit of small vector
boson mass. 
Our main result was presented in~\cite{Chachamis:2008xu}.
This was a second step towards the
complete evaluation of the
virtual corrections. In a forthcoming publication, we will derive a series
expansion in the mass and integrate the result numerically to recover the
full mass dependence, similarly to what has been done in~\cite{Czakon:2008zk}.

To complete the NNLO project one still needs to consider 
$2 \to 3$ real-virtual contributions  
and $2 \to 4$ real ones. The real-virtual corrections
are known from the NLO studies on $W W + jet$ production
in~\cite{Campbell:2007ev,
Dittmaier:2007th}.


\begin{thebibliography}{9}


\bibitem{tevatron1}
CDF Collaboration, Phys. Rev. Lett. {\bf 94} (2005) 211801

\bibitem{tevatron2}
  D0 Collaboration, Phys. Rev. Lett. 94 (2005) 151801

\bibitem{Spira:1995rr}
M.~Spira, A.~Djouadi, D.~Graudenz and P.~M.~Zerwas,
   Nucl.\ Phys.\  B {\bf 453} (1995) 17

\bibitem{Dawson:1990zj}
S.~Dawson,
   Nucl.\ Phys.\  B {\bf 359} (1991) 283

\bibitem{Harlander:2002wh}
R.~V.~Harlander and W.~B.~Kilgore,
   Phys.\ Rev.\ Lett.\  {\bf 88} (2002) 201801

\bibitem{Anastasiou:2002yz}
C.~Anastasiou and K.~Melnikov,
   Nucl.\ Phys.\  B {\bf 646} (2002) 220

\bibitem{Ravindran:2003um}
V.~Ravindran, J.~Smith and W.~L.~van Neerven,
   Nucl.\ Phys.\  B {\bf 665} (2003) 325

\bibitem{Catani:2001cr}
S.~Catani, D.~de Florian and M.~Grazzini,
   JHEP {\bf 0201} (2002) 015

\bibitem{Davatz:2004zg}
G.~Davatz, G.~Dissertori, M.~Dittmar, M.~Grazzini and F.~Pauss,
   JHEP {\bf 0405} (2004) 009

\bibitem{Anastasiou:2004xq}
C.~Anastasiou, K.~Melnikov and F.~Petriello,
   Phys.\ Rev.\ Lett.\  {\bf 93} (2004) 262002

\bibitem{Anastasiou:2007mz}
C.~Anastasiou, G.~Dissertori and F.~Stockli,
   arXiv:0707.2373 [hep-ph]

\bibitem{Grazzini:2008tf}
M.~Grazzini,
  arXiv: 0801.3232 [hep-ph]


\bibitem{Bredenstein:2006rh}
A.~Bredenstein, A.~Denner, S.~Dittmaier and M.~M.~Weber,
   Phys.\ Rev.\  D {\bf 74} (2006) 013004

\bibitem{kauer1}
T.~Binoth, M.~Ciccolini, N.~Kauer and M.~Kramer,
   JHEP {\bf 0503} (2005) 065  [arXiv:hep-ph/0503094]

\bibitem{kauer2}
T.~Binoth, M.~Ciccolini, N.~Kauer and M.~Kramer,
   JHEP {\bf 0612} (2006) 046  [arXiv:hep-ph/0611170]

\bibitem{dittmardreiner}
M.~Dittmar and H.~K.~Dreiner,
   Phys.\ Rev.\  D {\bf 55} (1997) 167 [arXiv:hep-ph/9608317]

\bibitem{brown}
R.~W.~Brown and K.~O.~Mikaelian,
   Phys.\ Rev.\  D {\bf 19} (1979) 922

\bibitem{ohn}
J.~Ohnemus,
   Phys.\ Rev.\  D {\bf 44} (1991) 1403

\bibitem{fri}
S.~Frixione,
   Nucl.\ Phys.\  B {\bf 410} (1993) 280

\bibitem{dixon1}
L.~J.~Dixon, Z.~Kunszt and A.~Signer,
   Nucl.\ Phys.\  B {\bf 531} (1998) 3  [arXiv:hep-ph/9803250]

\bibitem{dixon2}
L.~J.~Dixon, Z.~Kunszt and A.~Signer,
   Phys.\ Rev.\  D {\bf 60} (1999) 114037  [arXiv:hep-ph/9907305]

\bibitem{campbell}
J.~M.~Campbell and R.~K.~Ellis,
   Phys.\ Rev.\  D {\bf 60} (1999) 113006  [arXiv:hep-ph/9905386]

\bibitem{grazzini:2006}
M. Grazzini,
  JHEP {\bf 0601} (2006) 095

\bibitem{Anastasiou:2002zn}
C.~Anastasiou, E.~W.~N.~Glover and M.~E.~Tejeda-Yeomans,
   Nucl.\ Phys.\  B {\bf 629} (2002) 255


\bibitem{Chachamis:2007cy}
  G.~Chachamis,
  Acta Phys.\ Polon.\  B {\bf 38} (2007) 3563
  arXiv:0710.3035 [hep-ph]


\bibitem{Chachamis:2008yb}
G.~Chachamis, M.~Czakon and D.~Eiras,
  arXiv:0802.4028 [hep-ph]


\bibitem{Chachamis:2008fx}
  G.~Chachamis,
  arXiv:0807.0548 [hep-ph]


\bibitem{Chachamis:2008xu}
  G.~Chachamis, M.~Czakon and D.~Eiras,
  arXiv:0806.3043 [hep-ph]


\bibitem{qqTT}
M. Czakon, A. Mitov and S. Moch,
  Phys. Lett. B651 (2007) 147, arXiv:0705.1975 [hep-ph]

\bibitem{ggTT}
M.~Czakon, A.~Mitov and S.~Moch,
    arXiv:0707.4139 [hep-ph]

\bibitem{Czakon:2004wm}
M. Czakon, J. Gluza and T. Riemann,
  Phys. Rev. D71 (2005) 073009, hep-ph/0412164

\bibitem{Czakon:2006pa}
M. Czakon, J. Gluza and T. Riemann,
  Nucl. Phys. B751 (2006) 1, hep-ph/0604101

\bibitem{Actis:2007gi}
S.~Actis, M.~Czakon, J.~Gluza and T.~Riemann,
  Nucl.\ Phys.\  B {\bf 786} (2007) 26


\bibitem{Smirnov:1999gc}
V.A. Smirnov,
  Phys. Lett. B460 (1999) 397, hep-ph/9905323


\bibitem{Tausk:1999vh}
J.B. Tausk,
  Phys. Lett. B469 (1999) 225, hep-ph/9909506


\bibitem{MBrepresentation}
G. Chachamis and M. Czakon,
  {\tt MBrepresentation.m}, Unpublished


\bibitem{Czakon:2005rk}
M. Czakon,
  Comput. Phys. Commun. 175 (2006) 559, hep-ph/0511200


\bibitem{Moch:2005uc}
S. Moch and P. Uwer,
  Comput. Phys. Commun. 174 (2006) 759, math-ph/0508008


\bibitem{pslq:1992}
H.R.P. Ferguson and D.H. Bailey,
  (1992), (see e.g. http://mathworld.wolfram.com/PSLQ\\
Algorithm.html)


 


\bibitem{Diener:1997nx}
K.~P.~O.~Diener, B.~A.~Kniehl and A.~Pilaftsis,
  Phys.\ Rev.\  D {\bf 57} (1998) 2771

\bibitem{Denner:1988tv}
A.~Denner and T.~Sack,
  Nucl.\ Phys.\  B {\bf 306}, 221 (1988)


\bibitem{catani}
S.~Catani,
   Phys.\ Lett.\  B {\bf 427} (1998) 161  [arXiv:hep-ph/9802439].



\bibitem{Catani:1996vz}
S.~Catani and M.~H.~Seymour,
   Nucl.\ Phys.\  B {\bf 485} (1997) 291
   [Erratum-ibid.\  B {\bf 510} (1998) 503]


\bibitem{Czakon:2008zk}
M.~Czakon,
   arXiv:0803.1400 [hep-ph]



\bibitem{Campbell:2007ev}
J.~M.~Campbell, R.~K.~Ellis and G.~Zanderighi,
   JHEP {\bf 0712} (2007) 056

\bibitem{Dittmaier:2007th}
S.~Dittmaier, S.~Kallweit and P.~Uwer,
  Phys.\ Rev.\ Lett.\  {\bf 100} (2008) 062003




\end{thebibliography}
\end{document}